# On the (Im)possibility of Preserving Utility and Privacy in Personalized Social Recommendations

Ashwin Machanavajjhala
mvnak@yahoo-inc.com

Aleksandra Korolova
korolova@cs.stanford.edu

Atish Das Sarma
atish@cc.gatech.edu

## ABSTRACT

With the recent surge of social networks like Facebook, new forms of recommendations have become possible – personalized recommendations of ads, content, and even new social and product connections based on one's social interactions. In this paper, we study whether "social recommendations", or recommendations that utilize a user's social network, can be made without disclosing sensitive links between users. More precisely, we quantify the loss in utility when existing recommendation algorithms are modified to satisfy a strong notion of privacy called differential privacy. We propose lower bounds on the minimum loss in utility for any recommendation algorithm that is differentially private. We also propose two recommendation algorithms that satisfy differential privacy, analyze their performance in comparison to the lower bound, both analytically and experimentally, and show that good private social recommendations are feasible only for a few users in the social network or for a lenient setting of privacy parameters.

## 1. INTRODUCTION

Making recommendations or suggestions to users to increase their degree of engagement is a common practice for websites. For instance, Facebook recommends friends to existing users, Amazon suggests products, and Netflix recommends movies, in each case with the goal of making as *relevant* a recommendation to the user as possible. Recommending the right content, product, or ad to an individual is one of the most important tasks in today's web companies. With the boom in social networking many companies are striving to incorporate the likes and dislikes of an individual's social neighborhood. There has been much research and industrial activity to solve two problems: (a) recommending content, products, ads not only based on the individual's prior history but also based on the history of those the individual trusts [12, 2], and (b) recommending others whom the individual might trust. Recommendations based on social connections are especially effective for users who have seen very few movies, bought only a couple of products, or never clicked on ads; while traditional recommender systems default to generic recommendations, a social-network aware system can provide useful recommendations based on active friends. Companies like TrustedOpinion[1] and SoMR[2] generate content and ad recommendations by leveraging social networks. In fact, Facebook[3], Yahoo![4] and Google[5] are opening their social networks to third party developers to encourage social network-aware recommender systems.

In addition, a social network might want to use a different underlying social network, such as one derived from e-mail records or Instant Messenger connections, to suggest friends (e.g. Facebook already uses contacts imported from an address book as suggestions). Social connections could also be used to recommend products or advertisements to users— Netflix (or Opentable or Yelp) could recommend movies (or restaurants) to a subscriber based on her friends' activities and ratings. In fact, rather than using the entire social graph, the system could use only a subset of *trusted* edges for that application (for instance, a user might only trust the movie recommendations of a subset of her friends).

However, these improved recommendations based on social connections come at a cost – a recommendation can potentially lead to a *privacy breach* by revealing sensitive information. For instance, while the social network links might be public, both the user-product links and the user-user-trust links must be kept secret. (Knowing that your friend doesn't trust your judgement about books might be a breach of privacy). Similarly, revealing an edge in an e-mail graph, or revealing that a particular user purchased a sensitive product, constitutes a potentially serious breach of user privacy. Recommendations can indeed lead to such privacy breaches even without the use of social connections in the recommendation algorithm [5]. The privacy concerns posed by recommender systems and use of the social network graph have been at the forefront of industry discussion on the topic. In 2007, Facebook attempted to incorporate the product purchases made by one's friends into the stream of news one receives while visiting the site through a product called Beacon. Their launch showed that people interact with many websites and products in a way that they would not want their friends to know about, leading to several privacy lawsuits, and an eventual complete removal of the feature by Facebook.

---

[1] http://www.trustedopinion.com/
[2] http://www.somrnetworks.com/
[3] http://developers.facebook.com/connect.php
[4] http://developer.yahoo.com/yos/intro/
[5] http://www.google.com/friendconnect/

In this paper, we present the first theoretical study of the privacy-utility trade-offs in social recommender systems. While there are many different settings where social recommendations are used (friend/product recommendations, or trust propagation), and each leads to a slightly different formulation of the privacy problem (the sensitive information is different in each case), all these problems have the following common structure – recommendations are made on a graph where some subset of edges are sensitive. For clarity of exposition, we ignore (by and large) scenario specific constraints, and focus on the following general model. We consider a graph where all the edges are sensitive, and an algorithm that recommends a single node $v$ in the graph to some target node $u$. We assume that the algorithm is based on a utility function that encodes the "goodness" of recommending each node in the graph to this target node. Suggestions for utility functions include number of common neighbors, weighted paths and PageRank distributions [21]. We consider an attacker who wishes to deduce the existence of a single edge $(x,y)$ in the graph by passively observing the recommendation $(v,u)$. We measure the privacy of the algorithm using differential privacy – the ratio of the likelihoods of the algorithm recommending $(v,u)$ on the graphs with the edge $(x,y)$ and without the edge $(x,y)$, respectively. In this setting, we ask the question: to what extent can edge recommendations be accurate while preserving differential privacy?

**Our Contributions and Overview.** In this paper we present the following results on the accuracy of differentially private social recommendations.

- We present a trade-off between the accuracy and privacy of any social recommendation algorithm that is based on any general utility function. This trade-off shows an inevitable lower bound on the privacy parameter $\epsilon$ that must be incurred by an algorithm that wishes to guarantee any constant-factor approximation of the maximum utility. (Section 4)

- We present lower bounds on accuracy and privacy for algorithms based on specific utility functions previously suggested for recommending edges in a social network – number of common neighbors and weighted paths [21]. Our trade-offs for these specific utility functions present stronger lower bounds than the general one that is applicable for all utility functions. (Section 5)

- We adapt two well known privacy preserving algorithms from the differential privacy literature for the problem of social recommendations. The first (which we call Laplace), is based on adding random noise drawn from a Laplace distribution to the utility vector [8] and then recommending the highest utility node. The second (Exponential), is based on exponential smoothing [15]. We analyze and compare the accuracy of the two algorithms and comment on their relative merits. (Section 6)

- We perform experiments on a real graph using the number of common neighbors utility function. The experiments compare the algorithms Laplace, Exponential, and our lower bound. Our experiments suggest three takeaways: (i) For most nodes, the lower bounds suggest that there is a huge inevitable trade-off between privacy and accuracy when making social recommendations; (ii) The more natural Laplace algorithm performs as well as Exponential; and (iii) For a large fraction of nodes, both Laplace and Exponential almost achieve the maximum accuracy level suggested by our theoretical lower bound. (Section 7)

- We briefly consider the setting when an algorithm may not know (or be able to compute efficiently) the entire utility vector. We recognize that both Laplace and Exponential algorithms assume the knowledge of all the utilities (for every node) when recommending to a target node. We propose and analyze a sampling based linear smoothing algorithm that does not require all utilities to be pre-computed. We conclude by mentioning several directions for future work. (Section 8)

We now discuss related work and then formalize the models and definitions in Section 3.

## 2. RELATED WORK

Several papers propose that the social connections available can be effectively utilized for enhancing online applications [12, 2]. Golbeck [10] uses the trust relationships expressed through social connections for personalized movie recommendations and shows that the accuracy of the ratings outperform those produced by a collaborative filtering algorithm not utilizing the social graph. Mislove et al. [16] attempt an integration of web search with social networks and explore the use of trust relationships, such as social links, to thwart unwanted communication [17]. Approaches incorporating trust models into recommender systems are gaining momentum both in academic research [25], [18], [23], and in real products. Examples include, Chorus[6], which provides social app recommendations for the iPhone; Fruggo.com[7], a social e-commerce site; and WellNet's[8] online social networking program for health care coordination[9].

Calandrino et al. [5] demonstrate that algorithms that recommend products based on a friends' purchases have very practical privacy concerns: "passive observations of Amazon.com's recommendations are sufficient to make valid inferences about individuals' purchase histories". McSherry et al. [14] show how to adapt the leading algorithms used in the Netflix prize for movie recommendations to make privacy-preserving recommendations. Their work does not apply to algorithms that rely on the underlying social graph between users, as the user-user connections have not been released as part of the Netflix competition. Aïmeur et al. [1] study the problem of personalized recommendations in general. Dwork et al. [9] pose the problem of constructing differentially private analysis of social networks. Toubiana et al. [24] propose a framework for privacy preserving targeted advertising – while targeting based on user history is considered, targeting based on social interactions is not considered.

A related and independent work [4] considers the problem of mining top-k frequent item-set. Although they consider mechanisms analogous to the ones we propose, since we solve two different problems, the focus of their analysis, notion of utility, and conclusions substantially differ from ours.

---

[6] http://www.envionetworks.com/
[7] http://fruugo.com/
[8] http://www.wellnet.com/
[9] www.medicalnewstoday.com/articles/118628.php

## 3. MODEL

In this section, we describe the model for privacy-preserving social recommendations. We first define a social recommendation algorithm and then mention notions of monotonicity and accuracy of an algorithm. We then define axioms followed by typical utility functions that such algorithms are based on. Finally, we define differential privacy.

### 3.1 Social Recommendation Algorithm

Let $G = (V, E)$ be the graph that describes the social network. Each recommendation is an edge $(i, r)$, where node $i$ is recommended to the *target node* $r$. Given a graph $G$, and a target node $r$, we denote the utility of recommending node $i$ to node $r$ by $u_i^{G,r}$. Further, we assume that a recommendation algorithm $R$ is a probability vector on all nodes. Let $p_i^{G,r}$ denote the probability of recommending node $i$ to node $r$ in graph $G$ by a specified algorithm. When the graph $G$ and the source node $r$ are clear from context, we drop $G$ and $r$ from the notation – $u_i$ denotes the utility of recommending $i$, and $p_i$ denotes the probability that $R$ recommends $i$. We further define $u_{\max} = \max_i u_i$.

We consider algorithms that attempt to maximize the expected utility ($\sum_i u_i \cdot p_i$) of each recommendation. If we assume (without loss of generality) that the utility of the least useful recommendation is 0, the accuracy of such an algorithm can be defined as:

DEFINITION 1 (ACCURACY). *An algorithm $A$ is said to be $(1-\delta)$-accurate if given any set of utilities $u_i$ (for all $i$) denoted by $\vec{u}$, $A$ recommends node $i$ with probability $p_i$ such that $(1-\delta) = \min_{\vec{u}} \frac{\sum u_i p_i}{u_{\max}}$.*

Therefore, an algorithm is said to be $(1 - \delta)$-accurate if for any utility vector, the algorithm's expected utility is at least $(1-\delta)$ times the utility of the highest utility node in the given utility vector. It is easy to check that for the case when the utility of the least useful recommendation is $u_{\min}$, in all of our subsequent discussions, the definition of accuracy we use is equivalent to accuracy defined as the fraction of the difference between $u_{\max}$ and $u_{\min}$.

**Scale Invariance of Sensitivity and Utility Functions.** We initiate a small discussion on what happens when the utility values for all potential recommendable nodes are scaled by a multiplicative factor, or changed by an additive constant. Intuitively, since the scale of utilities is chosen arbitrarily, one would expect the algorithms and the analysis to be invariant to such numeric changes. However, because of the constraints imposed by the desire to be privacy-preserving, where the privacy-preservation is with respect to a presence or absence of a particular edge, the scale invariance assumptions require a more careful articulation. In particular, the crucial point of interaction between the privacy requirement and the utility function is the concept of sensitivity, denoted by $\Delta f$, which is the maximum change in a utility vector $\vec{u}$ that can occur due to an addition or removal of one edge in the graph. Observe that if we scale a utility function by a multiplicative constant, the sensitivity of the utility function is scaled as well by the same constant. Without loss of generality, and for ease of subsequent exposition, we assume that $\Delta f = 1$, an assumption that implies that the magnitudes of the utilities are now meaningful, as the higher utility magnitude corresponds to more edges that need to be added or removed in the graph in order to achieve it. Equivalently, we could have chosen to let the utilities be scale invariant, but would then need to compute and reason in terms of the sensitivity of the utility function.

Another property that is natural of a recommendation algorithm is monotonicity:

DEFINITION 2 (MONOTONICITY). *An algorithm is said to be monotonic if $\forall i, j$, $u_i \geq u_j$ implies that $p_i \geq p_j$.*

### 3.2 Axioms on Utility Functions

We now define two axioms that we believe should be satisfied by any meaningful utility function in the context of recommendations on a social network. These axioms are later used in proving our theoretical results. Our axioms are inspired by the work of [21] and the specific utility functions they consider, which include: number of common neighbors, sum of weighted paths, and PageRank based utility measures.

AXIOM 1 (EXCHANGEABILITY). *Let $G$ be a graph and let $h$ be an isomorphism on the nodes giving graph $G_h$, s.t. for target node $r$, $h(r) = r$. Then $\forall i : u_i^{G,r} = u_{h(i)}^{G_h,r}$.*

This axiom captures the intuition that the utility of a node $i$ should not depend on the node's name. Rather, its utility with respect to target node $r$ only depends on the structural properties of the graph, and so, nodes that are isomorphic from the perspective of the target node $r$ should have the same utility.

AXIOM 2 (CONCENTRATION AXIOM). *There exists $S \subset V(G)$, such that $|S| = \beta$, and $\sum_{i \in S} u_i \geq \Omega(1) \sum_{i \in V(G)} u_i$.*

This says that there are some $\beta$ nodes that together have at least a constant fraction of the total utility mass. This axiom is likely to be satisfied for small enough $\beta$, since usually there are some nodes that are very good for recommendation and many that are not so good.

In the subsequent lower bound sections, we only consider monotonic algorithms for utility functions that satisfy the exchangeability axiom as well as the concentration axiom for a reasonable choice of $\beta$.

A running example throughout the paper of a utility function that satisfies these axioms in practical settings and is often deployed [21] is that of the *number of common neighbors utility function*: given a target node $r$ and a graph $G$, the common neighbors utility metric assigns a utility $u_i^{G,r} = c(i,r)$, where $c(i,r)$ is the number of common neighbors between $i$ and $r$.

### 3.3 Differential privacy

Differential privacy [6] is a strong definition of privacy that is based on the following principle: an algorithm preserves the privacy of an entity if the algorithm's output is not sensitive to the presence or absence of the entity's information in the input data set. In our setting of graph-based social recommendations, we wish to maintain the presence (or absence) of an edge in the graph private. Hence, the privacy definition can be formally stated as follows.

DEFINITION 3. *A recommendation algorithm R satisfies $\epsilon$-differential privacy if for any pair of graphs G and G' that differ in one edge (i.e., $G = G' + \{e\}$ or vice versa) and every set of possible recommendations S,*

$$Pr[R(G) \in S] \leq exp(\epsilon) \times Pr[R(G') \in S] \quad (1)$$

*where the probabilities are over the random coins of R.*

Differential privacy has been widely used in the privacy literature [3, 8, 13, 15, 7], since it is even resilient to adversaries who know all but one edges in the graph, and guarantees privacy for multiple runs of the algorithm. While weaker notions of privacy have also been considered in the literature, in this paper we focus on the strong differential privacy definition only. Since in social recommendations protecting privacy is extremely important, it seems reasonable to first explore and understand the strongest notions of privacy.

In this paper, we only consider the utility of a *single* social recommendation. We note that in this setting, we can relax the differential privacy definition such that Equation 1 only holds for graphs $G$ and $G'$ that differ in an edge $e$ that is not incident on $r$, the target of the recommendation. This mirrors the natural setting where (a) one recommendation is made to the attacker ($r$), (b) only the target node (the attacker) sees the recommendation. By considering $G$ and $G'$ that differ in $e = (i, r)$, the adversary can only learn about his neighborhood (which he is aware of to start with) and not learn whether two legitimate nodes in the graph are connected. While we consider a single recommendation throughout the paper, we use the relaxed variant of differential privacy *only* in Sections 5 and 7.

## 4. GENERAL LOWER BOUND

In this section we prove a lower bound on the privacy parameter $\epsilon$ on any differentially private recommendation algorithm that (a) achieves a constant accuracy and (b) is based on any utility function that satisfies the exchangeability and concentration.

Let us first sketch the proof technique for the lower bound using the number of common neighbors utility metric, and then state the lower bound for a general utility metric. An interested reader can find the full proofs in the Appendix. Recall that given a target node $r$ and a graph $G$, the common neighbors utility metric assigns a utility $u_i^{G,r} = c(i, r)$, where $c(i, r)$ is the number of common neighbors between $i$ and $r$. The nodes in any graph can be split into two groups – $V_{hi}^r$, nodes which have a high utility for the target node $r$ and $V_{lo}^r$, nodes that have a low utility. In the case of common neighbors, all nodes $i$ in the 2-hop neighborhood of $r$ (who have at least one common neighbor with $r$) can be part of $V_{hi}^r$ and the rest in $V_{lo}^r$. Since the recommendation algorithm has to achieve a constant accuracy, it has to recommend one of the high utility nodes with constant probability.

By the concentration axiom, there are only a few nodes in $V_{hi}^r$, but there are many nodes in $V_{lo}^r$; in the case of common neighbors, node $r$ may only have 10s or 100s of 2-hop neighbors in a graph of millions of users. Hence, there exists a node $i$ in the high utility group and a node $\ell$ in the low utility group such that $\Gamma = p_i/p_\ell$ is very large ($\Omega(n)$). At this point, we show that we can carefully modify the graph $G$ by adding and/or deleting a small number ($t$) of edges in such a way that the node $\ell$ with the smallest probability of being recommended in $G$ becomes the node with the highest utility in $G'$. By the exchangeability axiom, we can show that there always exist some $t$ edges that make this possible. For instance in the common neighbors case, we can do this by adding edges between a node $i$ and $t$ of $r$'s neighbors, where $t > \max_i c(i, r)$. It now follows from differential privacy that

$$\epsilon \geq \frac{1}{t} \log \Gamma$$

More generally, let $V_{hi}^r$ be the set of nodes $1, \ldots, k$ each of which have utility $u_i > (1-c)u_{\max}$, and let $V_{lo}^r$ be the nodes $k+1, \ldots, n$ each of which have $u_i \leq (1-c)u_{\max}$ utility of being recommended to target node $r$. Recall that $u_{\max}$ is the utility of the highest utility node. Let $t$ be the number of edge alterations required to turn a node with the smallest probability of being recommended from the low utility group into a node of maximum utility in the modified graph. The following lemma states the main trade-off relationship between the accuracy parameter $\delta$ and the privacy parameter $\epsilon$ of a recommendation algorithm.

LEMMA 1. $\epsilon \geq \frac{1}{t}\left(\ln(\frac{c-\delta}{\delta}) + \ln(\frac{n-k}{k+1})\right)$

This lemma gives us a lower bound on the privacy guarantee $\epsilon$ in terms of the utility parameter $\delta$. Equivalently,

COROLLARY 1. $1 - \delta \leq 1 - \frac{c(n-k)}{n-k+(k+1)e^{\epsilon t}}$

By using the concentration axiom with parameter $\beta$ we can prove the following.

LEMMA 2. *For $(1-\delta) = \Omega(1)$ and $\beta = o(n/\log n)$,*

$$\epsilon \geq \frac{\log n - o(\log n)}{t} \quad (2)$$

This expression can be intuitively interpreted as follows: in order to achieve good accuracy with a reasonable amount of privacy (where $\epsilon$ is independent of $n$), either the number of nodes with high utility needs to be very large (i.e. $\beta$ needs to be very large, $\Omega(n/\log n)$), or the number of steps needed to bring up any node's utility to the highest utility needs to be large (i.e. $t$ needs to be large, $\Omega(\log n)$).

We shall use this relationship from Lemma 2 in the subsequent section to prove stronger lower bounds for specific utility functions. Below we mention a generic lower bound that applies to *any* utility function. Note that we only need an upper bound on $t$. The tighter upper bound we are able to prove on $t$, the better lower bound we get for $\epsilon$.

Using the exchangeability axiom, we can show that $t \leq 4 * d_{\max}$ in any graph. Consider the highest utility node and the lowest utility node, say $x$ and $y$ respectively. These nodes can be *interchanged* by deleting all of $x$'s current edges, adding edges from $x$ to $y$'s neighbors, and doing the same for $y$. This requires at most $4 * d_{\max}$ changes. Hence,

THEOREM 1. *For a graph with maximum degree $d_{\max} = \alpha \log n$, a differentially private algorithm can guarantee constant accuracy only if*

$$\epsilon \geq \frac{1}{\alpha}\left(\frac{1}{4} - o(1)\right) \quad (3)$$

In the next section, we present stronger lower bounds for two well studied utility functions – common neighbors and weighted paths.

# 5. LOWER BOUNDS FOR SPECIFIC UTILITY FUNCTIONS

In this section, we start from Lemma 2 and prove stronger lower bounds for specific utility functions by proving stronger upper bounds on $t$. Proofs and more details can be found in the Appendix.

## 5.1 Common neighbors lower bound

Consider a graph and a target node $r$. As we saw in the previous section, we can make any node $x$ have the highest utility by adding edges to all of $r$'s neighbors. If $d_r$ is $r$'s degree, it suffices to add $t = d_r + O(1)$ edges to make a node the highest utility node. We state the following theorem for a more general version of common neighbors utility function below.

THEOREM 2. *Let $U$ be a utility function that depends only on and is monotonically increasing with $c(x, y)$, the number of common neighbors between $x$ and $y$. A recommendation algorithm based on $U$ that guarantees any constant approximation to utility for target node $r$ has a lower bound on privacy given by $\epsilon \geq \frac{1-o(1)}{\alpha}$ where $d_r = \alpha \log n$.*

As we will show in Section 7, this is a very strong lower bound. Since a significant fraction of nodes in real-world graphs have small $d_r$ (due to a power law degree distribution), we can expect no algorithm based on common neighbors utility to be both accurate on most nodes and satisfy differential with a reasonable $\epsilon$.

## 5.2 Weighted Paths

A natural extension of the common neighbors utility function and one whose usefulness is supported by the literature [21], is the weighted path utility function, defined as

**score**$(s, y) = \sum_{l=2}^{\inf} \gamma^{l-1} |paths_{(s,y)}^{(l)}|$,

where $|paths_{(s,y)}^{(l)}|$ denotes the number of length $l$ paths from $s$ to $y$. Typically, one would consider using small values of $\gamma$, such as $\gamma = 0.005$, so that the weighted paths score is a "smoothed version" of the common neighbors score.

Again let $r$ be the target node. To make node $y$ the highest utility node, we add edges such that $y$ has $cd_r$ common neighbors with $r$. Now, the goal is to choose $c > 1$ such that this alone is sufficient to ensure that $y$ has the highest utility. This is done by showing that (a) no other node has more than $d_r$ common neighbors with $r$, and (b) the utility derived from paths of length $\geq 3$ cannot offset the additional common neighbors between $y$ and $r$ (for suitably small $\gamma$). Finally, we show that this requires adding only $t < d_r + 2*(c-1) + O(1)$.

THEOREM 3. *For weighted paths based utility functions with parameter $\gamma$, we have $t \leq (1+o(1))d_r$ when making recommendations for node $r$, if $\gamma = o(\frac{1}{d_{\max}})$. Therefore, for an algorithm to guarantee constant approximation to utility, the privacy must be $\epsilon \geq \frac{1}{\alpha}(1-o(1))$ where $d_r = \alpha \log n$.*

# 6. PRIVACY-PRESERVING RECOMMENDATION ALGORITHMS

There has been a wealth of literature on developing differentially private algorithms [3, 8, 15]. In this section we will adapt two well known tools, Laplace noise addition [8] and exponential smoothing [15], to our problem. For the purpose of this section, we will assume that given a graph and a target node, our algorithm has access to (or can efficiently compute) the utilities $u_i$ for all other nodes in the graph. Given this vector of utilities, our goal is to compute a vector of probabilities $p_i$ such that (a) $\sum_i u_i \cdot p_i$ is maximized, and (b) differential privacy is satisfied.

Clearly, maximum accuracy is achieved by recommending the node with utility $u_{max}$. However, it is well known that any algorithm that satisfies differential privacy must recommend every node, even the ones that have zero utility, with a non-zero probability [20]. Indeed, suppose for graph $G$ and target node $r$, an algorithm assigns 0 probability to some node $x$ with utility $u_x^{G,r}$ and a positive probability to some node $y$, with utility $u_y^{G,r}$. Transform $G$ into $G'$ as follows: connect $x$ to all of $y$'s neighbors in $G$ and disconnect $x$ from all its neighbors in $G$. Do the same for $y$. This in turns creates an isomorphism $h$ between $G$ and $G'$, where $h(r) = r$. Hence, by the exchangeability axiom, the algorithm will recommend $y$ with 0 probability. Thus, there is a path from $G$ to $G'$ of length $t$ such that $p_y$ goes from a positive number to 0. This leads to a breach of differential privacy.

The following two algorithms ensure differential privacy:

## 6.1 Exponential mechanism

The exponential mechanism creates a smooth probability distribution from the utility vector and then samples from that.

DEFINITION 4. **Exponential mechanism:** *Given nodes with utilities $(u_1, \ldots, u_i, \ldots, u_n)$, algorithm $A_E(\epsilon)$ recommends node $i$ with probability*
$e^{\frac{\epsilon}{\Delta f} u_i} / \sum_{k=1}^{n} e^{\frac{\epsilon}{\Delta f} u_k}$, *where $\epsilon \geq 0$ is the privacy parameter, and $\Delta f$ is the sensitivity of the utility function*[10].

THEOREM 4. *[15] $A_E(\epsilon)$ guarantees $\epsilon$ differential privacy.*

## 6.2 Laplace mechanism

Unlike the exponential mechanism, the Laplace mechanism mimics the optimal mechanism. It first adds random noise drawn from a Laplace distribution, and like the optimal mechanism, picks the node with the maximum noise-infused utility.

DEFINITION 5. **Laplace mechanism:** *Given nodes with utilities $(u_1, \ldots, u_i, \ldots, u_n)$, algorithm $A_L(\epsilon)$ first computes a modified utility vector $(u'_1, \ldots, u'_n)$ as follows: $u'_i = u_i + r$ where $r$ is a random variable chosen from the Laplace distribution with scale $(\frac{\Delta f}{\epsilon})$[11] independently at random for each $i$. Then, $A_L(\epsilon)$ recommends node $z$ whose noisy utility is maximal among all nodes, i.e. $z = \arg\max_i u'_i$.*

THEOREM 5. *$A_L(\epsilon)$ guarantees $\epsilon$ differential privacy.*

PROOF. The proof follows from the privacy proof of the Laplace mechanism in the context of publishing privacy-preserving histograms [8] by observing that one could treat each node as a histogram bin and release the noisy count for the value in that bin, $u'_i$. Since $A_L(\epsilon)$ is effectively doing post-processing by releasing only the name of the bin with the highest noisy count, the algorithm remains private. □

---

[10] $\Delta f = \max_r \max_{G,G':G=G'+e} ||u^{\vec{G},r} - u^{\vec{G'},r}||$
[11] In this distribution, the pdf at $y$ is $\frac{\epsilon}{2\Delta f} \exp(-|y|\epsilon/\Delta f)$

An astute reader might remark at this point that the Laplace mechanism as stated does not satisfy the monotonicity property that we relied upon in our lower bound proofs. Indeed, the Laplace mechanism satisfies the property only in expectation; however, that is not an obstacle to our analysis since in order to meaningfully compare the performance of Laplace mechanism with other mechanisms and with the theoretical bound on performance, we would need to evaluate its expected, rather than one-time, performance.

### 6.3 Exponential vs Laplace Mechanisms

It is natural to ask whether there is an equivalence between the two approaches of transforming a non-private algorithm to a privacy-preserving algorithm or how they would compare, perhaps depending on the setting. We present preliminary results on comparing the utilities when there are only two possible recommendations ($n = 2$). The theorem is stated below and the proof can be found in the Appendix.

THEOREM 6. *Let $U_E$ and $U_L$ denote the utilities achieved by $A_E(\epsilon)$ and $A_L(\epsilon)$ on input vector $(u_1, u_2)$, respectively. Wlog, assume $u_1 \geq u_2$. Then $U_E = u_1 \frac{e^{\epsilon u_1}}{e^{\epsilon u_1} + e^{\epsilon u_2}} + u_2 \frac{e^{\epsilon u_2}}{e^{\epsilon u_1} + e^{\epsilon u_2}}$ and*
$U_L = u_1(1 - \frac{1}{2}e^{-\epsilon(u_1-u_2)} - \frac{\epsilon(u_1-u_2)}{4e^{\epsilon(u_1-u_2)}}) + u_2(\frac{1}{2}e^{-\epsilon(u_1-u_2)} + \frac{\epsilon(u_1-u_2)}{4e^{\epsilon(u_1-u_2)}})$

To our knowledge, in the course of the proof we give the first explicit closed form expression for the probabilities of each of the two nodes being recommended by Laplace mechanism (the work of [19] gives a formula that does not apply to our setting).

Although the expressions for $U_E$ and $U_L$ are difficult to compare by eye-balling, by plugging in various values of $u_1$ and $u_2$ into the formulas, one infers that the Exponential mechanism slightly outperforms the Laplace mechanism, when $\epsilon$ is very small and the difference between $u_1$ and $u_2$ is large. We leave it for future work to simplify these as well as extend the analysis to the $n > 2$ case.

**Implementation efficiency.** The Laplace mechanism is more intuitive than the Exponential mechanism, and more likely to receive executive buy-in in a real-world environment. Furthermore, it has the advantage that it can be implemented more easily than the Exponential mechanism. $A_L$ requires computing the noisy utilities and then selecting the node with the highest noisy utility, which takes linear time. $A_E$ requires first computing a set of smoothed utilities and then sampling from the probability distribution induced by them, which can be accomplished in linear time using the alias-urn method suggested by [22], but likely slightly less practically efficiently than $A_L$.

## 7. UTILITY ACHIEVABLE IN PRACTICE ON A REAL GRAPH

In this section we present experimental results on a real graph and for the # of common neighbors utility function. The experiments compare the algorithms Laplace, Exponential, and our lower bound. Our experiments suggest three takeaways: (i) For most nodes, the lower bounds suggest that there is a huge inevitable trade-off between privacy and accuracy when making social recommendations; (ii) The more natural Laplace mechanism performs as well as the Exponential mechanism; and (iii) For a large fraction of nodes, the accuracy achieved by Laplace and Exponential mechanisms does not substantially differ from the best possible accuracy suggested by our theoretical lower bound.

### 7.1 Experimental Setup

For our experiments we use the Wikipedia vote network [11] available from Stanford Network Analysis Package[12]. Some users in Wikipedia are administrators, who have access to additional technical features. Users are elected to be administrators via a public vote of other users and administrators. The Wikipedia vote network consists of all users participating in the elections (either casting a vote or being vote on), since inception of Wikipedia until January 2008. We turn the network of [11] into an undirected network, where each node represents a user and an edge from node $i$ to node $j$ represents that user $i$ voted on user $j$ or user $j$ voted on user $i$. This obtained network consists of 7,115 nodes and 100,762 edges. Although the Wikipedia vote network is publicly available, and hence the edges in it are not private, we believe that the graph itself exhibits the structure and properties of some of the graphs in which one would want to preserve privacy, such as the graph of social connections and people's product purchases.

For each pair of nodes in the social network, except nodes that share an edge, we compute the number of common neighbors they have in the Wikipedia vote network. Then, assuming we will make one recommendation for each node in the graph, we compute the expected accuracy of recommendation for that node. For the Exponential mechanism and the theoretical bound, given the utilities of recommending each node to a given node $v$, we can compute the expected accuracy and the theoretical bound on accuracy exactly. For the Laplace mechanism, we compute its expected accuracy by running 1,000 independent trials of the Laplace mechanism, and averaging the utilities obtained in those trials, for each node in the graph.[13]

### 7.2 Exponential vs Laplace in practice

We first observe in Figure 1 that for all nodes in the Wikipedia vote network, the Laplace mechanism achieves nearly identical accuracy as the Exponential mechanism. This confirms our hypothesis of Section 6 that Exponential and Laplace mechanisms are nearly equivalent in practical settings, and implies that one can use the more intuitive and easily implementable Laplace mechanism in practice.

### 7.3 Social Recommendations: Good or Private?

We now proceed to evaluate the accuracy of the Exponential mechanism and compare it with the best accuracy one can hope to achieve using a privacy-preserving recommendation algorithm, as computed according to our theoretical bound of Corollary 1.

For ease of visual presentation, we assume that we do not care about node identities; we number the nodes in decreasing order of the accuracy one can hope for when making the recommendation for that node, as predicted by the theoretical bound. For each node, the graph in Figure 2 shows the theoretical bound and the accuracy achieved by the Ex-ponential mechanism. Due to our chosen numbering of the nodes, the theoretical bound is a smooth monotonically decreasing function of the node number, whereas the achieved accuracy is not necessarily monotonically decreasing (and thus, in places, does not appear as a line).

---

[12] http://snap.stanford.edu/data/wiki-Vote.html

[13] Out of the 7,115 nodes, there are 60 nodes that have no common neighbors with anyone except nodes they are already connected to. We omit those nodes from our analysis.

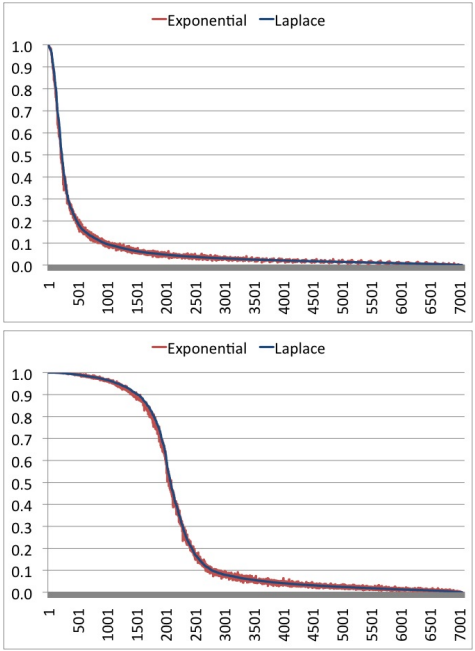

**Figure 1:** Accuracy achieved by Exponential and Laplace mechanisms on Wikipedia vote network using # of common neighbors as a measure of utility. The x-axis represents the node number, the y-axis - the accuracy of recommendation for that node. The top graph is for desired privacy guarantee of $\epsilon = 0.1$, the bottom - for $\epsilon = 0.5$.

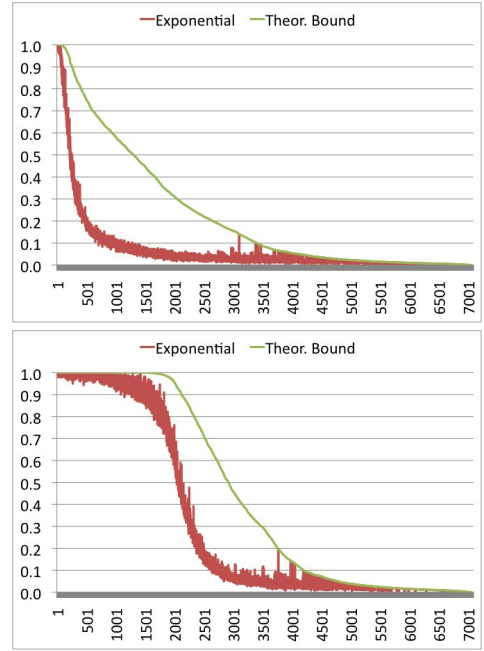

**Figure 2:** Accuracy achieved by Exponential mechanisms and predicted by theoretical bound on Wikipedia vote network using # of common neighbors as a measure of utility. The x-axis represents the node number, the y-axis - the expected accuracy of recommendation for that node. The top graph is for desired privacy guarantee of $\epsilon = 0.1$, the bottom - for $\epsilon = 0.5$.

As can be seen in Figure 2 and Figure 3, for some nodes, the Exponential mechanism performs quite well, achieving nearly perfect accuracy. However, the number of such nodes is fairly small - the Exponential mechanism achieves better than 0.9 approximation for less than 1.5% of the nodes when $\epsilon = 0.1$ and less than 21% of the nodes when $\epsilon = 0.5$, it achieves better than 0.8 approximation for less than 2% of the nodes when $\epsilon = 0.1$ and less than 25.5% of the nodes when $\epsilon = 0.5$. This matches the intuition that by making the privacy requirement more lenient, one can hope to make better quality recommendations for more nodes; however, this also pinpoints the fact that for most nodes, the Exponential mechanism does not achieve good accuracy.

Although there is a possibility that one could develop better privacy-preserving recommendation mechanisms than Exponential or Laplace, this experiment shows that for a large number of target nodes, our theoretical bound limits the best accuracy one can hope to achieve privately quite severely. For example, for $\epsilon = 0.1$, no privacy-preserving algorithm can hope to achieve a better than 70% accuracy for more than 9% of the nodes. This finding throws into serious question the feasibility of developing social recommendation algorithms that are both accurate and privacy-preserving for many real-world settings.

Finally, in practice, it is the least connected nodes that are likely to benefit most from receiving high quality recommendations. However, our experiments suggest that the low degree nodes are also the most vulnerable to receiving low accuracy recommendations due to needs of privacy-preservation: see Figure 4 for an illustration of how accuracy depends on the degree of the node. This further suggests that, in practice, one has to make a choice between preserving accuracy vs preserving privacy.

### 7.4 Are $A_E$ and $A_L$ good enough for utility function based on common neighbors?

As we have experimentally observed in Figure 2, the Exponential mechanism achieves good accuracy compared to the best achievable accuracy predicted by our theoretical bound. We can formalize this statement rigorously as follows (proved in the Appendix):

LEMMA 3. *Let $A_E$ denote the accuracy of the Exponential mechanism, and $A_O$ denote the upper bound on the accuracy that can be achieved by any privacy-preserving algorithm. Then, for utility functions based on the number of common neighbors between two nodes, $\frac{A_E}{A_O} \geq \frac{1}{k+1}$, where $k$ is the number of nodes with non-zero utility.*

Furthermore,

LEMMA 4. *For utility vector of the form $u = (u_{\max}, \ldots, u_{\max}, 0, \ldots, 0)$, $\frac{A_E}{A_O} \geq \frac{k}{k+1}$, where $k$ is the number of nodes with non-zero utility,*

For real-world graphs, we expect the number of nodes with non-zero utility $k$ to be fairly small, and thus, the Expo-nential mechanism to achieve a good approximation to the best possible accuracy achievable by a privacy-preserving social recommendation algorithm. Furthermore, observe that Corollary 1 merely gives an upper bound on accuracy achievable in a privacy-preserving manner, but it might be the case that tighter lower bounds can be obtained. Hence, in many ways, the Exponential and Laplace mechanisms are representative of the class of good privacy-preserving mechanisms one can hope for.

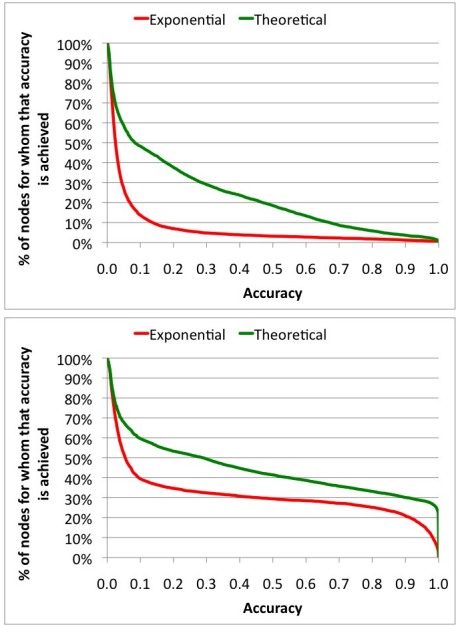

**Figure 3: Performance of the Exponential mechanisms and predicted by theoretical bound on Wikipedia vote network using # of common neighbors as a measure of utility. The x-axis represents the accuracy, the y-axis - the expected percentage of nodes for whom that accuracy of recommendation is achieved (or predicted by the theoretical bound). The top graph is for desired privacy guarantee of $\epsilon = 0.1$, the bottom - for $\epsilon = 0.5$.**

# 8. EXTENSIONS AND FUTURE WORK

## 8.1 Vertex privacy and non-monotone algorithms

We considered the setting of graph based social recommendations where we wished to maintain private the information about the presence or absence of an edge in the graph but our reasoning and results can easily be generalized to a setting where we would like to protect the entire identity of a node. To achieve that, one would need to strengthen the definition of the recommendation algorithm satisfying differential privacy to consider graphs that differ in one node, rather than one edge, and adjust the value of $t$, the number of edge alterations to turn a node from the low utility group into a node of maximum utility, respectively.

Furthermore, our results can be generalized to social recommendation algorithms that do not satisfy the monotonicity property. For clarity of exposition, we omit the exact statements and proofs of lemmas analogous to Lemmas 1 and 2 but remark that the statement formulations and our qualitative conclusions will remain essentially unchanged, with the exception of the meaning of variable $t$. Without the monotonicity property, $t$ would correspond to the number of edge alterations necessary to *exchange* the node with the smallest probability of being recommended and the node with the highest utility, rather than to the number of edge alterations necessary to *make* the node with the smallest probability of being recommended into the node with the highest utility, leading to a higher value for $t$.

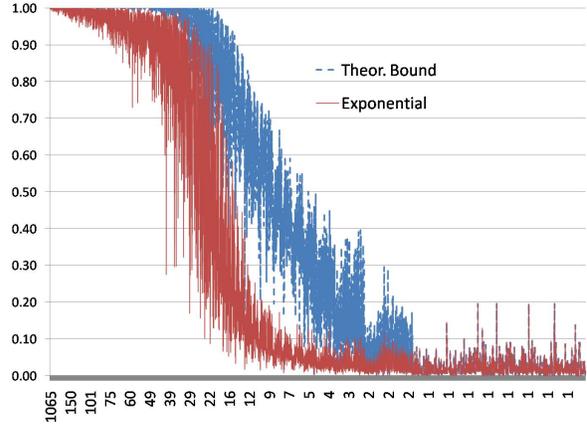

**Figure 4: Accuracy achieved by Exponential mechanism and predicted by Theoretical Bound as a function of node degree, $\epsilon = 0.5$**

## 8.2 What if utility vectors are unknown?

Both the differentially private algorithms we considered in Section 6 assume the knowledge of the entire utility vector. This assumption cannot be made in social networks for various reasons. Firstly, computing as well as storing the utility of $n^2$ pairs is prohibitively expensive, when dealing with graphs of several hundred million nodes. Secondly, even if one could compute and store them, these graphs change at staggering rates, therefore, utility vectors are also constantly changing. We believe that this is a very important and interesting problem. In this section, we explore a simple algorithm that assumes no knowledge of the utility vector; it only assumes that sampling from the utility vector can be done efficiently.

### 8.2.1 Sampling and Linear Smoothing

Suppose we are given an algorithm $A$ which is a $\gamma$ approximation in terms of utility, and not provably private. We show how to modify the algorithm $A$ to guarantee differential privacy, while still preserving, to some extent, the utility approximation of $A$. The proof of the following theorem, and a note, are placed in the appendix.

DEFINITION 6. *Given algorithm $A = (p_1, \ldots, p_i, \ldots, p_n)$, algorithm $A_S(x)$ recommends node $i$ with probability $\frac{1-x}{n} + xp_i$, where $0 \leq x \leq 1$ is a parameter.*

THEOREM 7. *$A_S(x)$ guarantees $\ln(1 + \frac{nx}{1-x})$-differential privacy and a $x\gamma$ approximation of utility.*

Another idea worth exploring is perturbing the input graph (by adding/deleting a fraction of possible edges) and then sampling and recommending from it. What is the relationship between the extent of perturbation and the utility/privacy guarantees?

### 8.3 Future Work

Several interesting questions remain unexplored in this work. While we have considered some specific utility functions in this paper, it would be nice to look more. Further, our motivation was to look at the most stringent requirement in terms of privacy; however, a natural question is to understand utility-privacy trade-offs for certain typical graphs that arise in social networks.

This paper only considers lower bounds and algorithms for making one single recommendation. It would be very interesting, and important, to explore how the effect on privacy compounds with multiple recommendations. Further, some edges can be more sensitive than others. Perhaps the solution should be methodological - enable opt-in/opt-out settings to specify which nodes/edges are private. A closer look at such dependences is required.

Also, most works on making recommendations deal with static databases. Social networks clearly change over time (and rather rapidly). This raises several issues, such as not being able to assume the utility vector is known, sensitivity changing, privacy impacts of dynamic databases etc. Dealing with such temporal graphs and understanding there trade-offs would be very interesting.

Finally, it would certainly be interesting to extend these results for weaker notions of privacy than differential privacy. For instance, some privacy notions previously defined include $k$-anonymity, $(\epsilon, \delta)$-differential privacy, and relaxing the adversary's background knowledge to just the general statistics of the graph.

## 9. ACKNOWLEDGMENTS

The authors are grateful to Arpita Ghosh and Tim Roughgarden for thought-provoking discussions.

# APPENDIX

**Proof of Lemma 1**

PROOF. We initiate the analysis with a simple claim.

CLAIM 1. *In order to achieve $(1-\delta)$ accuracy, at least $\frac{c-\delta}{c}$ of the probability weight has to go to nodes in the high utility group, so there exists a node $x$ in the low utility group of $G_1$ that is recommended with probability of at most $\frac{\delta}{c(n-k)}$, e.g. $p_x^{G_1} \leq \frac{\delta}{c(n-k)}$.*

PROOF. Denote by $p^+$ and $p^-$ the total probability that goes to high/low utility nodes, respectively, and observe that $p^+ u_{\max} + (1-c)u_{\max}p^- \geq \sum_i u_i p_i \geq (1-\delta)u_{\max}$ and $p^+ + p^- \leq 1$, hence, $p^+ > \frac{c-\delta}{c}, p^- \leq \frac{\delta}{c}$. □

We now continue the proof of Lemma 1.

Let $G_2$ be the graph that turns $x$, found according to the Claim above, into a node of highest utility by addition of $t$ edges.

By differential privacy, we have $\frac{p_x^{G_2}}{p_x^{G_1}} \leq e^{\epsilon t}$.

In order to achieve $(1-\delta)$ accuracy on $G_2$, at least $\frac{c-\delta}{c}$ of the probability weight has to go to nodes in the high utility group, and hence by monotonicity $Pr[x|G_2] > \frac{c-\delta}{c(k+1)}$. Combining the previous three inequalities, we obtain:
$\frac{(c-\delta)(n-k)}{(k+1)\delta} = \frac{\frac{c-\delta}{c(k+1)}}{\frac{\delta}{c(n-k)}} < \frac{p_x^{G_2}}{p_x^{G_1}} \leq e^{\epsilon t}$, hence

$$\epsilon \geq \frac{1}{t}\left(\ln(\frac{c-\delta}{\delta}) + \ln(\frac{n-k}{k+1})\right)$$

This completes the proof. □

**Proof of Lemma 2**

PROOF. We first use the concentration axiom to prove the following claim.

CLAIM 2. *If $c = \left(1 - \frac{1}{\log n}\right)$, then $k = O(\beta \log n)$ where $\beta$ is the parameter of the concentration axiom.*

PROOF. Now consider the case when $c = \left(1 - \frac{1}{\log n}\right)$.

Therefore, $k$ is the number of nodes that have utility at least $\frac{u_{\max}}{\log n}$. Let the total utility mass be $U = \sum_i u_i$. Since by concentration, the $\beta$ highest utility nodes add up to a total utility mass of $\Omega(1) * U$, we have $u_{\max} \geq \Omega(\frac{U}{\beta})$. Therefore, $k$, the number of nodes with utility at least $\frac{u_{\max}}{\log n}$ is at most $\frac{U \log n}{u_{\max}}$ which is at most $O(\beta \log n)$. □

We now prove the Lemma using Lemma 1 and Claim 2.

Substituting these in the expression, if we need $1 - \frac{c(n-k)}{n-k+(k+1)e^{\epsilon t}}$ to be $\Omega(1)$, then require $(k+1)e^{\epsilon t}$ to be $\Omega(n-k)$. (Notice that if $(k+1)e^{\epsilon t} = o(n-k)$, then $\frac{c(n-k)}{n-k+(k+1)e^{\epsilon t}} \geq c - o(1)$, which is $1 - o(1)$.).

Therefore, if we want an algorithm to obtain constant approximation in utility, i.e. $(1-\delta) = \Omega(1)$, then we need the following (assuming $\beta$ to be small):

$$(O(\beta \log n))e^{\epsilon t} = \Omega((n - O(\beta \log n))$$

Or (for small enough $\beta$)

$$e^{\epsilon t} = \Omega(\frac{n}{\beta \log n})$$

Simplifying

$$\epsilon \geq \frac{\log n - \log \beta - \log \log n}{t}$$

$$\epsilon \geq \frac{\log n - o(\log n)}{t}$$

□

**Proof of Theorem 2**

PROOF. **Lower Bound for Common Neighbors** We formalize the intuition in terms of an upper bound on $t$ in the following claim.

CLAIM 3. *For common neighbors based utility functions, when recommendations for $r$ are being made, we have $t \leq d_r + 2$, where $d_r$ is the degree of node $r$.*

PROOF. Observe that if the number of common neighbors is the measure of the utility of recommendation, then one can make any zero utility node, say $x$, for source node $r$ into a max utility node by adding $d_r$ edges to all of $r$'s neighbors and additionally adding two more edges (one each from $r$ and $x$) to some node with small utility. This is because the highest utility node has at most $d_r$ common neighbors with $r$ (one of which could potentially be $x$). Further, adding these edges cannot increase the number of common neighbors for any other node beyond $d_r$. □

We now use this to get the theorem immediately by replacing $t$ in the expression stated previously. □

**Proof of Theorem 3**

PROOF. **Lower Bound for Sum of Weighted Paths**

The number of paths of length $l$ between two nodes is at most $d_{\max}^{l-1}$. Let $x$ be the highest utility node and let $y$ be the node we wish to make the highest utility node after adding certain edges. If we are making recommendations for node $r$, then the maximum number of common neighbors with $r$ is at most $d_r$.

Currently denote the utility of $x$ by $u_x$. We know that $u_x \leq \gamma d_r \sum_{l=3}^{\inf} \gamma^{l-1} d_{\max}^{l-1}$. (In fact one can tighten the second term as well.)

We rewire the graph as follows. Any $(c-1)d_r$ nodes (other than $y$ and the source node $r$) are picked; here $c > 1$ is to be determined later. Both $r$ and $y$ are connected to these $(c-1)d_r$ nodes. Additionally, $y$ is connected to all of $r$'s $d_r$ neighbors. Therefore, we now get the following.

$$u_y \geq \gamma c d_r$$

Now we wish to bound by above the utility of any other node in the network in this rewired graph. Notice that every other node still has at most $d_r$ paths of length 2 with the source. Further, there are only two nodes in the graph that have degree more than $d_{\max} + 1$, and they have degree at most $(c+1)d_{\max}$. Therefore, the number of paths of length $l$ for $l \geq 3$ for any node is at most $((c+1)d_{\max})^2 \cdot (d_{\max}+1)^{l-3}$. This can be further tightened to $((c+1)d_{\max})^2 \cdot (d_{\max})^{l-3}$. We therefore get the following for any $x$ in the rewired graph,

$$u_x \leq \gamma d_r + (c+1)^2 \sum_{l=3}^{\infty} \gamma^{l-1} d_{\max}^{l-1}$$

Now consider the case where $\gamma < \frac{1}{d_{\max}}$. We get

$$u_x \leq \gamma d_r + \frac{(c+1)^2 \gamma^2 d_{\max}^2}{1 - \gamma d_{\max}}$$

We now want $u_y \geq u_x$. This reduces to

$$(c-1) \geq \frac{(c+1)^2 \gamma d_{\max}}{1 - \gamma d_{\max}}$$

Now if $\gamma = o(\frac{1}{d_{\max}})$ then it is sufficient to have $(c-1) = \Omega(\gamma d_{\max})$ which can be achieved even with $c = 1 + o(1)$. Now notice that we only added $d_r + 2(c-1)d_r$ edges to the graph. This completes the proof of the theorem. □

**Comment on relationship between common neighbors and weighted paths:** Since common neighbors is an extreme case of weighted paths (as $\gamma \to 0$), we are able to obtain the same lower bound (up to $o(1)$ terms) when $\gamma$ is made small (in particular, $\gamma \approx o(\frac{1}{d_{\max}})$). Can one obtain (perhaps weaker) lower bounds when say $\gamma = \Theta(\frac{1}{d_{\max}})$? Notice that the proof only needs $(c-1) \geq \frac{(c+1)^2 \gamma d_{\max}}{1-\gamma d_{\max}}$. We then get a lower bound of $\epsilon \geq \frac{1}{\alpha}(\frac{1-o(1)}{2c-1})$ where $d_r = \alpha \log n$. Setting $\gamma d_{\max} = s$, for some constant $s$, we can find the smallest $c$ that satisfies the expression $(c-1) \geq \frac{(c+1)^2 s}{1-s}$. Notice that this does give a nontrivial lower bound (i.e. a lower bound tighter than the generic one presented in the previous section), as long as $s$ is a sufficiently small constant.

**Proof of Theorem 6**

PROOF. **Utility of Laplace for $n = 2$:** Suppose we have two elements, with utility $t_1$ and $t_2$, respectively, where $t_1 \geq t_2$ wlog.

Let $\phi_X(t)$ denote the characteristic function of the Laplace distribution, it is known that $\phi_X(t) = \frac{1}{1+b^2t^2}$. Moreover, it is known that if $X_1$ and $X_2$ are independently distributed random variables, then $\phi_{X_1+X_2}(t) = \phi_{X_1}(t)\phi_{X_2}(t) = \frac{1}{(1+b^2t^2)^2}$. Using the inversion formula, we can compute the pdf of $X = X_1 + X_2$ as follows:

$$f_X(x) = F'_X(x) = \frac{1}{2\pi} \int_{-\infty}^{\infty} e^{-itx} \phi_X(t) dt$$

For $x > 0$, the pdf of $X_1 + X_2$ is $f_X(x) = \frac{1}{4b}(1+\frac{x}{b})e^{-\frac{x}{b}}$ and the cdf is $F_X(x) = 1 - \frac{1}{4}\epsilon e^{-\epsilon x}(\frac{2}{\epsilon} + x)$.

What is the probability that element 1 is recommended? It's the $Pr[t_1 + X_1 > t_2 + X_2] = Pr[X_2 - X_1 < t_1 - t_2] = 1 - \frac{1}{4}\epsilon e^{-\epsilon(t_1-t_2)}(\frac{2}{\epsilon} + (t_1-t_2)) = 1 - \frac{1}{2}e^{-\epsilon(t_1-t_2)} - \frac{\epsilon(t_1-t_2)}{4e^{\epsilon(t_1-t_2)}}$

Hence, the Laplace mechanism recommends node 1 with probability

$$1 - \frac{1}{2}e^{-\epsilon(t_1-t_2)} - \frac{\epsilon(t_1-t_2)}{4e^{\epsilon(t_1-t_2)}},$$

from which the desired statement about $A_L$'s utility follows. □

**Proof of Theorem 7**

PROOF. **Sampling and Linear Smoothing**

Let $p''_i = \frac{1-x}{n} + xp_i$. We have

$$\frac{1-x}{n} \leq p''_i \leq \frac{1-x}{n} + x,$$

since $0 \leq p_i \leq 1$.

The utility of $A_S$ is

$$U(A_S) = \sum_{k=1}^{n} u_k p''_k = \sum_{k=1}^{n} (\frac{1-x}{n}) u_k + \sum_{k=1}^{n} xp_k u_k = \frac{1-x}{n} + x\gamma \geq x\gamma,$$

where we use $\sum_k u_k = 1$ and $\sum p_k u_k = \gamma$.

For the privacy guarantee, note again that the upper and lower bounds on $p'_i$ hold for *any* graph and utility function. Therefore, the change in the probability of recommending $i$ for any two graphs $G$ and $G'$ that differ in exactly one edge is at most

$$\frac{p_i(G)}{p_i(G')} \leq \frac{x + \frac{1-x}{n}}{\frac{1-x}{n}} = 1 + \frac{nx}{1-x}.$$

Therefore, $A_S$ is $\ln(1+\frac{nx}{1-x})$-differentially private. This complete the proof.

Further, note, to guarantee $2\epsilon$- differentially privacy for $A_S(x)$, we need to set the parameter $x$ so that $\ln(1+\frac{nx}{1-x}) = 2c \ln n$ (rewriting $\epsilon = c \ln n$), namely

$$x = \frac{n^{2c} - 1}{n^{2c} - 1 + n}.$$

The algorithm $A_S$ guarantees a utility of at least $x\gamma$. □

**Proof of Lemma 3**

PROOF. Suppose the variations on the common neighbor functions permitted are $u_i = d_i/z$, where $d_i$ is the number of common neighbors node $i$ has with the target node, and $z$ is a scaling constant. Pick $c = 1$, meaning that all nodes except $k$ have zero utility. Then $U_O \leq u_{\max}(1 - \delta) \leq u_{\max}(1 - \frac{n-k}{n-k+(k+1)e^{\epsilon u_{\max}}}) = u_{\max} \frac{(k+1)e^{\epsilon u_{\max}}}{n-k+(k+1)e^{\epsilon u_{\max}}}$.

Under our restricted privacy definition, the sensitivity of the scaled number of common neighbors utility function is $\frac{1}{z}$. $U_E \geq u_{\max} \frac{e^{\epsilon z u_{\max}}}{n-k+ke^{\epsilon z u_{\max}}} = u_{\max} \frac{e^{\epsilon u_{\max}}}{n-k+ke^{\epsilon u_{\max}}} \geq u_{\max} \frac{e^{\epsilon u_{\max}}}{n-k+(k+1)e^{\epsilon u_{\max}}}$. Hence $\frac{U_E}{U_O} \geq \frac{1}{k+1}$ and exponential algorithm gives a $(k+1)$ approximation of utility, which could be a fairly good approximation, if $k$ is small compared to $n$, which is what we expect, in real-world social network graphs. □